\documentclass[reprint, secnumarabic, amssymb, nobibnotes, aps, prl, floats, floatfix, nofootinbib]{revtex4-1}

\usepackage{graphicx}
\usepackage[export]{adjustbox}
\usepackage{amsmath}
\usepackage{color}

\begin{document}
\title{Full and unbiased solution of the Dyson-Schwinger equation in the functional integro-differential representation}

\author{Tobias Pfeffer}
\author {Lode Pollet}
\affiliation{Department of Physics, Arnold Sommerfeld Center for Theoretical Physics,
University of Munich, Theresienstrasse 37, 80333 Munich, Germany}
\date{\today}%

\begin{abstract}
We provide a full and unbiased solution to the Dyson-Schwinger equation illustrated for $\phi^4$ theory in 2D. It is based on an exact treatment of the functional derivative $\partial \Gamma / \partial G$ of the 4-point vertex function $\Gamma$ with respect to the 2-point correlation function $G$ within the framework of the homotopy analysis method (HAM) and the Monte Carlo sampling of rooted tree diagrams.
The resulting series solution in deformations can be considered as an asymptotic series around $G=0$ in a HAM control parameter $c_0G$, or even a convergent one up to the phase transition point if shifts in $G$ can be performed (such as by summing up all ladder diagrams).
These considerations are equally applicable to fermionic quantum field theories and offer a fresh approach to solving integro-differential equations.
\end{abstract}

\maketitle

%
%
%
Despite decades of research there continues to be a need for developing novel methods for strongly correlated systems. The standard Monte Carlo approaches \cite{RecentDevelopmentsPIMC, SSESandvik, AuxField, Determinants, ContTime} are convergent but suffer from a prohibitive sign problem, scaling exponentially in the system volume \cite{SignProblem}. Diagrammatic Monte Carlo simulations \cite{HappyPolaron, FermiPolaron, LecNotes} were developed to prevent this, scaling exponentially only in the expansion order \cite{PseudoPolComplex}. However, this happens at the expense of substantially worse series convergence properties \cite{FermiPolaron2DVlietinck, FermiPolaron2DKroiss, PseudoGap}. After nearly ten years and despite recent and tremendous progress \cite{ConnectedDeterminants, InchWorm, Keldysh}, one may well fear that the combination of an asymptotic/divergent series, even with a mild sign problem, is as prohibitive as the standard approaches. Recently~\cite{HAMPhi4truncated}, we therefore suggested to use the more flexible Dyson-Schwinger equation (DSE) instead of self-consistent Feynman diagrams \cite{BoldDiagMonteCarlo} to provide a fully self-consistent scheme on the one and two particle level. Furthermore, we extended the homotopy analysis method (HAM)~\cite{HAMbook1,HAMbook2} to $\phi^4$ field theory in two dimensions (2D) (providing us with more tools to enhance the convergence properties in a systematic way), and showed how the expansion in terms of rooted trees is amenable to a systematic Monte Carlo sampling. This expansion is furthermore convenient when dealing with multi-dimensional objects such as the 4-point vertex function. Clearly, this is a radically different way at looking at interacting field theories. However, in our previous work~\cite{HAMPhi4truncated} we truncated the DSE at the level of the 6-point vertex. The infinite tower of equations for $n$-point correlation functions was not solved and differences with the full, exact answer could be seen when the correlation length increases.

In this Letter we solve the full DSE by writing them as a closed set of integro-differential equations. Within the HAM theory there exists a semi-analytic way to treat the functional derivatives without resorting to an infinite expansion of the 
successive $n$-point correlation functions, cf. \cite{Phi4Buividovich, NonAbelianBuividovich} where the DSE has been used to generate new weak coupling expansions. The unbiased numerical solution of the DSE is largely unexplored as it was considered to be too complex to be solved even in the simplest cases \cite{DSETruncI, DSETruncII, DSEQCD}. Furthermore, taking into account the functional derivatives deteriorates the convergence properties of the field theory substantially compared to the truncated case considered in Ref.~\cite{HAMPhi4truncated}. Here we show how the remaining theory can be brought under control within the HAM as an asymptotic expansion of the HAM deformations in terms of an auxiliary convergence control parameter $c_0$ (times the 2-point correlation function $G$) around $G=0$, or even as a convergent expansion in the HAM deformations when a shift of $G$ is possible, e.g. by solving the ladder equations. Although the ideas are illustrated for a 1d integral and $\phi^4$ theory in 2D the convergence considerations and the methodology are generically applicable as long as the $2$-point and $4$-point correlation functions are bounded, and may just as well be applied to the better known Hedin \cite{Hedin} and parquet equations \cite{Parquet} or to the functional renormalization group equation \cite{fRGStatMech,fRGFermion}. Below, we first analyze the toy model (i.e. the 1d integral, cf. \cite{Regularization0D}) to illustrate the approach and then proceed with the full solution of the DSE for the $\phi^4$ model in 2D.\\

%
%
%
{\it Functional closure} -- The functional closure is most easily demonstrated in the case of a 0D field theory. Consider the 1D integral
\begin{equation}
\begin{split}
Z[J] = \int \!\! \text{d}\phi \; e^{-S[\phi] + J\phi} \text{ with } S[\phi] = \frac{1}{2} k \phi^2  + \frac{\lambda}{4!} \phi^4,
\end{split}
\label{Integral0D}
\end{equation}
with $Z[J]$ the generating functional of the $n$-point correlation functions $ G^{(n)} $,
\begin{equation}
 G^{(n)} = \frac{1}{Z[0]} \int \!\! \text{d}\phi \; e^{-S[\phi]} \phi^n = \frac{1}{Z[0]} \left. \frac{\text{d}^n Z[J]}{\text{d} J^n} \right|_{J=0}.
\end{equation}
Although in this example the generating functional $Z[J]$ is a real-valued function and the $n$-point correlation functions are real numbers, we will, nevertheless, use the terminology of (quantum) field theory (FT) as there will be no ambiguities. Moreover, we use the shorthand notation $G = G^{(2)}$ for the 2-point correlation function. The DSE can be derived by introducing an infinitesimal shift $\delta$ in the integration variable, $\phi \rightarrow \phi + \delta$, and expanding the resulting expression in powers of $\delta$. This yields
\begin{equation}
\frac{ \text{d} S }{ \text{d} \phi} \left[ \phi = \frac{ \text{d} }{ \text{d} J }\right] Z[J] = J Z[J].
\label{Diffeqn0D}
\end{equation}
The first derivative of the action $S$ which respect to the field $\phi$ is promoted to a differential operator by the substitution $\phi = \frac{ \text{d} }{ \text{d} J}$. The DSE (\ref{Diffeqn0D}) is a definition of the generating functional in terms of a differential equation equivalent to the definition of $Z[J]$ through (\ref{Integral0D}). For a realistic FT, (\ref{Diffeqn0D}) turns into a functional integro-differential equation, whereas (\ref{Integral0D}) turns into a functional integral.\\
Instead of focusing on the solution of (\ref{Integral0D}) we focus in the following on (\ref{Diffeqn0D}). Differentiating (\ref{Diffeqn0D}) once with respect to $J$ and setting $J=0$ afterwards yields, after introducing the connected 4-point correlation function $G^{(4)}_c = G^{(4)} - 3 G^2$ and the 4-point vertex function $\Gamma = G^{-4} G^{(4)}_c$, 
\begin{equation}
G^{-1} - k = \frac{\lambda}{2} G + \frac{\lambda}{6} G^3 \Gamma.
\label{fullclosedDSE0d_1}
\end{equation}
This is the first equation in the expansion of the differential equation (\ref{Diffeqn0D}) into an infinite hierarchy of coupled equations for the correlation functions.  We close the hierarchy by considering the generating functional to be a functional of the inverse bare 2-point correlation function $k$, $Z=Z[J=0,k]$.
By applying the chain rule $\frac{\text{d}}{\text{d} k} = \frac{\text{d} G}{\text{d} k} \frac{\text{d}}{\text{d} G}$ and using (\ref{fullclosedDSE0d_1}), we arrive at 
\begin{eqnarray}
\Gamma[G] =  &-\lambda - \frac{3\lambda}{2} G^2 \Gamma - \frac{\lambda}{2} G^4 \Gamma^2 \label{fullclosedDSE0d_2} \\ 
& - \frac{\lambda}{3} G^3 \Gamma^{\prime } -\frac{\lambda}{6} G^5 \Gamma^{\prime } \Gamma. \nonumber
\end{eqnarray}
We denote the derivative as $\Gamma^{\prime } = \frac{ \text{d} \Gamma[G]}{ \text{d} G} $. 
A solution to (\ref{fullclosedDSE0d_1}) requires knowledge of the universal functional $\Gamma[G]$ by solving the differential equation (\ref{fullclosedDSE0d_2}). $\Gamma[G]$ is subsequently used in (\ref{fullclosedDSE0d_1}) and solved for $G$ for a given inverse of the bare 2-point correlation function $k$. The combined system (\ref{fullclosedDSE0d_1}), (\ref{fullclosedDSE0d_2}) is typically solved by a fixed point iteration. This gives the physical 2-point correlation function $G$. The physical 4-point vertex function $\Gamma$ follows by evaluating the universal functional $\Gamma[G]$ for the physical $G$.
In Ref.~\onlinecite{HAMPhi4truncated} we have already shown how to solve a realistic FT when only taking the first three terms on the right hand side of (\ref{fullclosedDSE0d_2}) into account i.e. working with a truncated version of $\Gamma[G]$ which has a finite convergence radius. In the following we examine the deterioration in convergence properties when taking the derivatives in (\ref{fullclosedDSE0d_2}) into account. \\
\begin{figure}[t]
\centering
\includegraphics[width=0.9\linewidth]{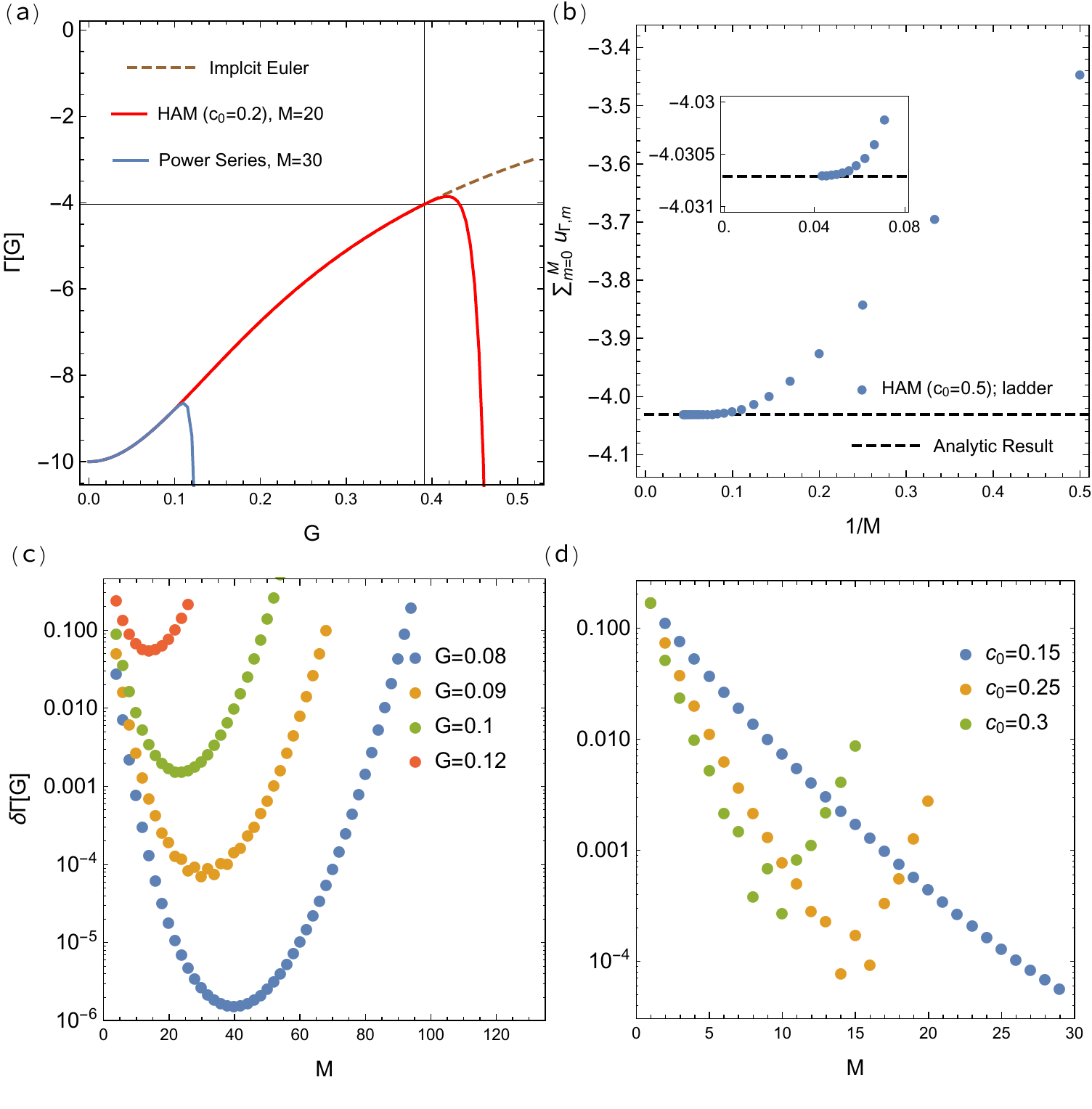}
\caption{(a) $\Gamma[G]$ at coupling $\lambda=10$ obtained by the implicit Euler algorithm, a power series (asymptotic), and the HAM method (asymptotic). The thin grid lines show the evaluation for the physical $G$, $\Gamma$ for $k=1$. (b) The convergence of the HAM series solution with respect to a linear operator $\mathcal{L}_{\text{ladder}}[\Gamma[G]] = \Gamma[G] + \lambda + \frac{3 \lambda}{2} G^2 \Gamma[G]$, the ladder approximation to the vertex equation (\ref{fullclosedDSE0d_2}). The functional $\Gamma[G]$ is evaluated at the physical $G$ for $k=1$. The inset shows the convergence of the main plot on a much finer scale.  (c) The relative error $\delta \Gamma[G]$ of a power series solution at small $G$ shows the behavior of an asymptotic series. (d) The relative error $\delta \Gamma[G]$ of the HAM series solution equivalent to a generalized Taylor series for $G=0.3$ shows the behavior of an asymptotic series controlled by the convergence control parameter $c_0$.}
\label{fig:FunctionalGammaG}
\end{figure}
%
%

{\it Vertex equation} -- 
The solution of (\ref{fullclosedDSE0d_2}), obtained by the implicit Euler method is shown in Fig.\,\ref{fig:FunctionalGammaG}(a) for $\lambda = 10$. It agrees with inverting the functional $G[k] \rightarrow k[G]$ and plugging it into $\Gamma[k[G]]$. The additional vertical (horizontal) grid line corresponds to the physical $G$($\Gamma$) obtained for the model parameters $k=1$, $\lambda = 10$.  
In case of a FT, (\ref{fullclosedDSE0d_2}) takes the form of a functional integro-differential equation and the implicit Euler method cannot be applied. It is therefore necessary to study semi-analytic solutions to the differential equation (\ref{fullclosedDSE0d_2}). The power series expansion $\Gamma[G] = -\lambda + \lim_{M\rightarrow\infty} \sum_{n = 1}^{M} c_n G^n$ has zero convergence radius~\cite{ComplexDE,SM}: Bringing (\ref{fullclosedDSE0d_2}) into the form of $ \Gamma^{\prime}[G] = F\left[ G, \Gamma^{\prime }[G] \right]$ we find that $F$ is not holomorphic at $G=0$. An expansion around $G=0$ corresponds to approaching the non-interacting limit by fixing $\lambda$ and taking $k \rightarrow \infty$, i.e. $Gk = \left[ \int \!\! \text{d}\phi \phi^2 \! \exp(- \frac{1}{2} \phi^2 - \frac{\lambda}{4!k^2} \phi^4 )  \right]/ Z \stackrel{k \to \infty}{=} 1$. We find numerically, see Fig.\,\ref{fig:FunctionalGammaG}(c), that the power series solution to (\ref{fullclosedDSE0d_2}) has the properties of an asymptotic expansion, {\it i.e.}, for every small $G$ there exists an optimal truncation order $M$ such that the truncated power series asymptotically approaches the exact answer with exponential accuracy. However, Fig.\,\ref{fig:FunctionalGammaG}(a) shows that it is not possible to construct $\Gamma[G]$ as a power series at values of $G$ where it corresponds to the physical $G$ for the considered model parameters $k = 1$, $\lambda = 10$.\\

A more powerful semi-analytic method is the HAM~\cite{HAMbook1}. The starting point is the construction of the homotopy 
\begin{equation}
(1-q)\mathcal{L} \left[ \Phi[G,q] - u_{\Gamma,0}[G] \right] + q c_0 \mathcal{N}\left[ \Phi[G,q] \right] = 0
\label{homotopy}
\end{equation}
for the differential equation (\ref{fullclosedDSE0d_2}). $\mathcal{N}\left[ \Gamma[G] \right] = 0$ is the non-linear differential operator defining (\ref{fullclosedDSE0d_2}) and $\mathcal{L}$ is an arbitrary linear operator with the property $\mathcal{L}[0]=0$. The homotopy (\ref{homotopy}) includes the deformation parameter $q  \in [0,1]$ which deforms the solution of $\mathcal{L}$ from $\Phi[G,0] = u_{\Gamma,0}[G]$ at $q=0$ to the solution of the differential equation (\ref{fullclosedDSE0d_2}), $\Phi[G,1] = \Gamma[G]$, at $q=1$. $u_{\Gamma,0}[G]$ is the initial guess for the solution of $\mathcal{N}\left[ \Gamma[G] \right] = 0$. The convergence control parameter $c_0$ controls the rate at which the deformation happens. The HAM attempts to find the solution of (\ref{homotopy}) through a Taylor series expansion in $q$, {\it i.e.}, $\Phi[G,q] = u_{\Gamma,0}[G] + \sum_{m=1} u_{\Gamma,m}[G] \; q^m$. The expansion coefficients are given by $u_{\Gamma,m}[G] = \frac{1}{m!} \Phi^{(m)}[G,q=0]$ and can be obtained by the $m$-th derivative of (\ref{homotopy}). Therefore, the HAM gives a series solution of the differential equation (\ref{fullclosedDSE0d_2}) in terms of the deformation coefficients $u_{\Gamma,m}[G]$, $\Gamma[G] = u_{\Gamma,0}[G] + \sum_{m=1} u_{\Gamma,m}[G]$. We first use the easiest possible linear operator $\mathcal{L}\left[ \Phi[G,q] - u_{\Gamma,0}[G] \right] = \Phi[G,q] - u_{\Gamma,0}[G]$. This choice can be straightforwardly generalized to functional integro-differential equations. It is well known that for this form of the linear operator  the series solution obtained by the HAM corresponds to a generalized Taylor series expansion, which enlarges the convergence radius of the ordinary Taylor series through the convergence control parameter $c_0$ \cite{HAMbook1}. 
Unfortunately, in order for the generalized Taylor theorem to hold, {\it i.e.}, that the convergence radius can be enlarged by choosing appropriate parameters $c_0$, the ordinary Taylor series must have a finite convergence radius. As we have argued before, the ordinary Taylor series has zero convergence radius, {\it i.e.}, the series solution of the HAM does not give a convergent answer for any $G$. 
The HAM convergence parameter $c_0$ gives us however additional freedom since the effective parameter $c_0G$ can always be made small as long as $G$ remains finite. As is illustrated in Figs.~\ref{fig:FunctionalGammaG}(a) and (d), the asymptotic nature can then be postponed to larger expansion orders for smaller $c_0$ but with larger deviations from the exact result at low expansion orders. This approach can be generalized to realistic field theories in which case $c_0 ||G||$ can be chosen to be small where $||G||$ denotes the $L^p$-norm of the 2-point correlation function. In the next section we show numerically that this holds for a generic FT and that it is possible to get accurate results even close to a 2nd order phase-transition at which $||G||$ gets unbounded.
Before extending this approach to a realistic FT we show how to obtain a convergent HAM series solution. Instead of constructing a generalized Taylor series around the non-interacting limit $G = 0$ we use the linear operator $\mathcal{L} = \mathcal{L}_{\text{ladder}}$ to construct a homotopy around the analytically solvable ladder approximation $\mathcal{L}_{\text{ladder}}[\Gamma[G]] = \Gamma[G] + \lambda + \frac{3 \lambda}{2} G^2 \Gamma[G]$. With this linear operator the HAM series solution is convergent as shown in Fig.~\ref{fig:FunctionalGammaG}(b). Moreover, as the HAM series solution is in this case not equivalent to the generalized Taylor series we find a globally convergent solution, i.e. it is not restricted by a finite convergence radius. For further details see Ref.~ \onlinecite{SM}.\\

%
%
{\it $\phi^4$ theory in 2d} -- 
\begin{figure}[t]
\centering
\includegraphics[width=0.9\linewidth]{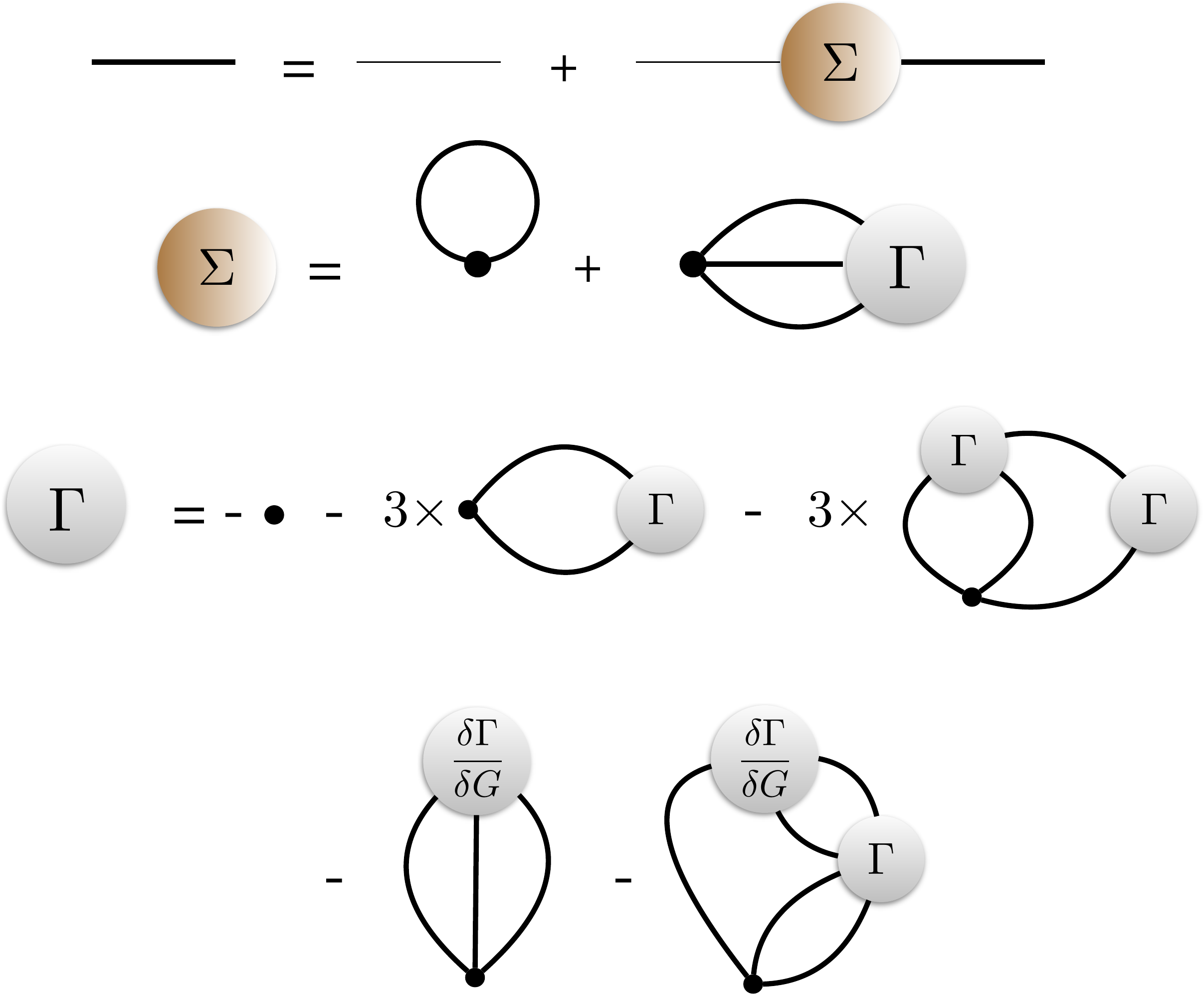}
\caption{The coupled set of equations defining model (\ref{Action2D}) through a functional integro-differential equation. Each diagram is in one-to-one correspondence with the terms in (\ref{fullclosedDSE0d_1}), (\ref{fullclosedDSE0d_2}). 
A bare propagator $G_{0;i,j}$ is denoted by a thin line, the 2-point correlation function $G_{i,j}$ by a bold line, and the bare vertex by a dot. The correct convolution of lattice indices can be obtained by standard diagrammatic rules. The terms without functional derivatives involve the permutation of external indices and is denoted by the factor $3$.}
\label{fig:FunctionalIntegroEqn}
\end{figure}
Consider the $\mathcal{Z}_2$-symmetric $\phi^4$-theory on a 2D lattice with action
\begin{equation}
S[\phi] = \frac{1}{2} \sum_{i,j} \phi_i G^{-1}_{0;i,j} \phi_j + \frac{\lambda}{4!} \phi_i^4.
\label{Action2D}
\end{equation}
The inverse bare 2-point correlation is given by $G^{-1}_{0;i,j} = -\Box_{i,j} + \frac{1}{2} m^2  \delta_{i,j}$ where $\Box_{i,j}$ denotes the discretized Laplace operator in 2D. For $m^2<0$ this model undergoes a second order phase transition from a magnetically ordered to an unordered phase at a critical coupling constant $\lambda_c(m^2)$. The phase transition is signalled by the divergence of the magnetic susceptibility $\chi = G(p=0)$ which should lead to the same critical exponents as for the 2D Ising model previously studied diagrammatically with grassmannization techniques \cite{GrassmanizationIsing}. In our simulation we use $m^2 = -0.5$ which corresponds to $\lambda_c(m^2) \sim 2$. The coupled set of Eqs.~(\ref{fullclosedDSE0d_1}) and (\ref{fullclosedDSE0d_2}) are graphically depicted in Fig.\,\ref{fig:FunctionalIntegroEqn}. 

\begin{figure}[t]
\centering
\includegraphics[width=0.9\linewidth]{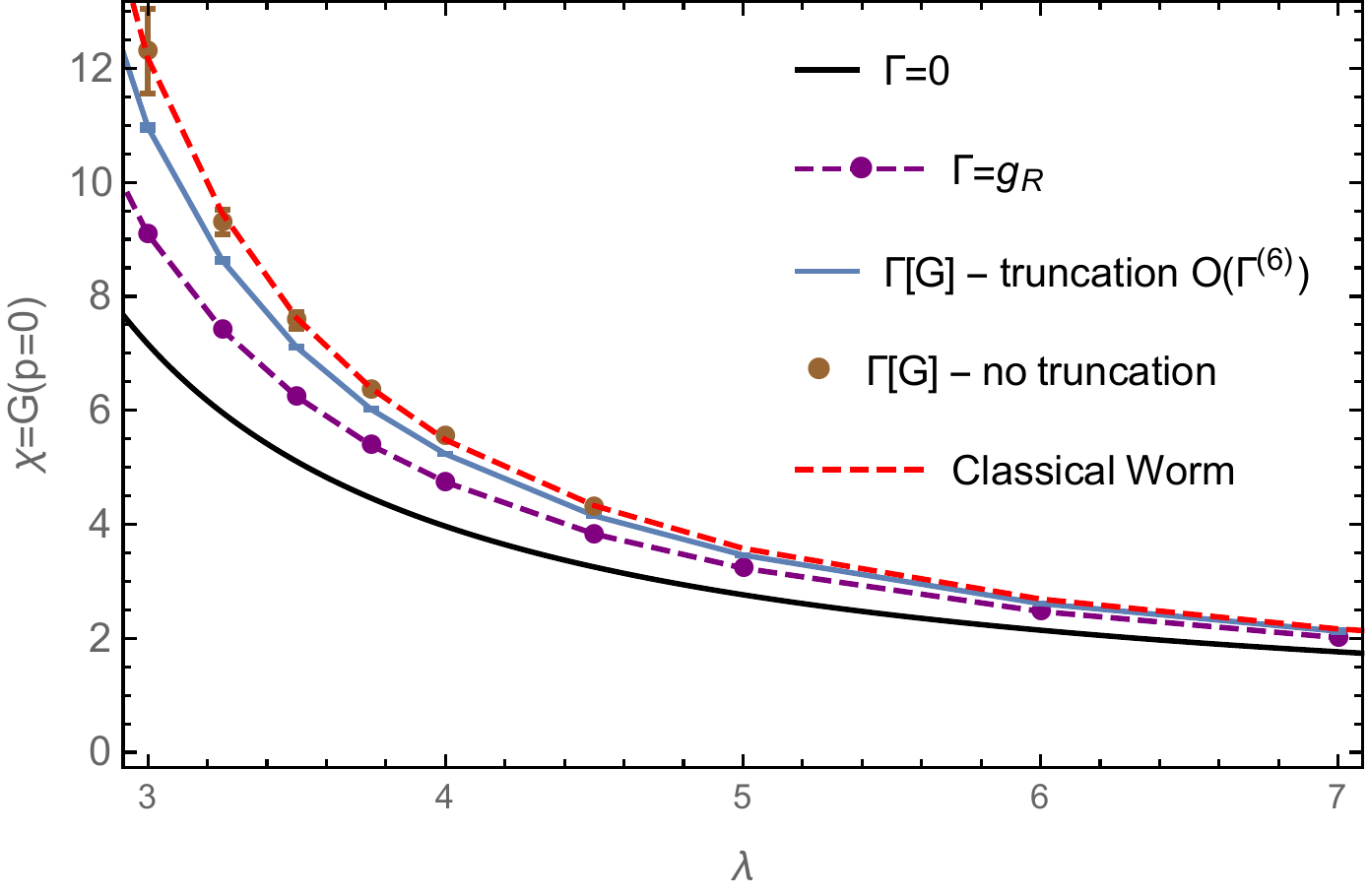}
\caption{The development of the divergence of the susceptibility $\chi = G(p=0)$ close to the phase transition for the model (\ref{Action2D}). Various approximations of the universal functional $\Gamma[G]$ yield systematic, quantitative errors whereas the full solution captures the correct quantitative behavior. The black, dashed purple and blue lines were obtained in Ref.~\onlinecite{HAMPhi4truncated} and correspond to different truncations of the functional $\Gamma[G]$. The black approximation corresponds to $\Gamma[G] = 0$, the dashed purple line to $\Gamma[G] = -\frac{\lambda}{1+\frac{3 \lambda}{2} \int_p\! G(p)^2}$ and the blue line to the truncation of $\Gamma[G]$ which takes in account only the first three diagram elements on the right hand side of Fig.~\ref{fig:FunctionalIntegroEqn} in the functional integro-differential equation for $\Gamma[G]$ i.e. if corrections of the order of the 6-point vertex function $\Gamma^{(6)}$ are neglected.}
\label{fig:ResultSusc}
\end{figure}

\begin{figure}[t]
\centering
\includegraphics[width=0.9\linewidth]{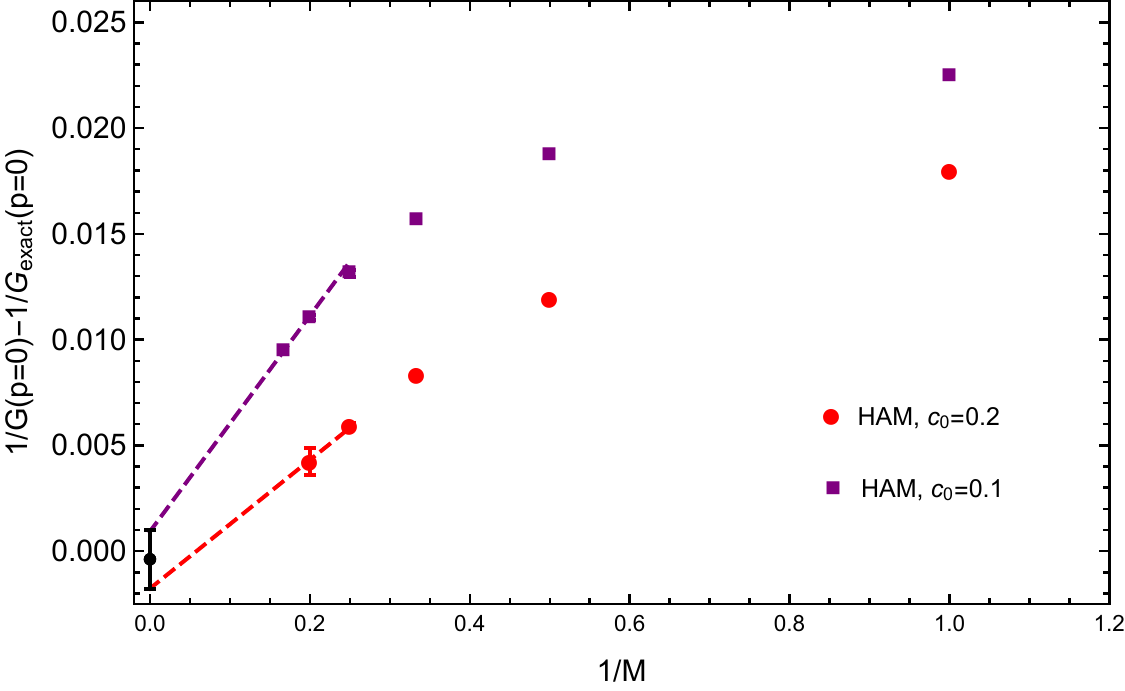}
\caption{In order to obtain an error estimate for the extrapolation of the homotopy series we use a linear extrapolation for different deformation parameters $c_0$. We make sure within errorbars that results extrapolate to the same value for different $c_0$. The plot shows the results for $\lambda = 3.5$ where the correlation length is already large enough such that there are considerable deviation from the exact result if only a truncated functional $\Gamma[G]$ has been considered, cf. Fig.~\ref{fig:ResultSusc}.}
\label{fig:Extrapolation}
\end{figure}

We use an extension of the Monte Carlo algorithm developed in Ref.~\onlinecite{HAMPhi4truncated} to sample the expansion of the HAM in rooted tree diagrams. We will describe the detailed algorithm, especially the correct implementation of the functional derivatives, for an arbitrary many-body system with 2-body interactions elsewhere.
Fig.~\ref{fig:ResultSusc} shows the result for the divergence of the susceptibility $\chi$ for successive approximations of $\Gamma[G]$ and for the full solution of the functional integro-differential equation. The results are compared to the numerically exact simulation of model Eq.~(\ref{fig:FunctionalIntegroEqn}) with the classical Worm algorithm \cite{ClassicalWorm} on system sizes which are much larger than the correlation length.
We find that it is possible to obtain controlled results with our current Monte Carlo algorithm up to $\chi \sim 10$ which corresponds to a correlation length of $\xi \sim 3$. Our diagrammatic Monte Carlo sampling is based on a direct sampling of all topologies of rooted tree diagrams at a given order. This naive sampling leads to sampling problems at higher deformation orders and restricts the order to 5-6. We find that the sign problem is of the same order as for the stochastic construction of $\Gamma[G]$ in Ref.~\onlinecite{HAMPhi4truncated}. We performed extensive Monte Carlo simulations such that the error bars are exclusively determined by the extrapolation of the first deformation orders as is shown in Fig.\,\ref{fig:Extrapolation}. It is only possible to extrapolate the inverse of the 2-point correlation function and therefore the error introduced through the extrapolation is growing rapidly as the susceptibility diverges, cf. Fig.~\ref{fig:ResultSusc}.
It should be pointed out that we have no global access to $\Gamma[G]$ but only to a stochastic, local evaluation of the universal functional through the diagrammatic Monte Carlo sampling of the HAM series solution in terms of rooted tree diagrams. In order to obtain the results in Fig.\,\ref{fig:ResultSusc} through this evaluation of $\Gamma[G]$ we simplified the calculation by starting the fixed point iteration for the coupled system (\ref{fullclosedDSE0d_1}), (\ref{fullclosedDSE0d_2}) at the numerically exact 2-point correlation function. This reduces the computational time as we have to perform only a single fixed point iteration step. For more general starting points of the fixed point iteration we refer to Ref.~\onlinecite{HAMPhi4truncated}.
Figs.~\ref{fig:ResultSusc} and \ref{fig:Extrapolation} show that, although there is no single physical small parameter close to the phase transition, we obtain controlled results close to the transition as we can construct the homotopy (\ref{homotopy}) always with respect to a small enough $c_0$. In the vicinity of the second order phase transition,  $||G|| \rightarrow \infty$ implies that $c_0 \rightarrow 0$ but in order to get meaningful results, the number of required deformations increases rapidly. The sign problem is then the limiting factor. For transitions where $G$ remains finite (the divergence may then occur in the 4-point vertex function), this argument is not applicable.\\

{\it Outlook} -- Although the approach to FT introduced in this Letter is illustrated for the simple though representative case of $\phi^4$-theory, models with arbitrary 2-body interactions in its symmetric phase can be tackled on the same footing. The reformulation of the functional integral representation for general models to functional integro-differential equations yields exactly the same equation as depicted in Fig.~\ref{fig:FunctionalIntegroEqn}. The only difference is that the convolution of indices in the diagrams representing the functional integro-differential equation is with respect to a collective index $i$ which summarizes all possible field labels of the considered model. The statistics of fermionic fields translate into additional signs for the permutations of external indices in the diagrams of Fig.~\ref{fig:FunctionalIntegroEqn} and into sign alternating 2-point correlation functions.\\

{\it Conclusion} -- In conclusion, we have introduced a general approach to tackle FT through full and unbiased solutions of functional integro-differential equations derived from the DSE. We showed for a toy model that by using the semi-analytic HAM we can solve the differential equation for the vertex function. The HAM gives a convergent series solution if the homotopy is constructed with respect to the analytically solvable ladder approximation for the 4-point vertex function or an asymptotic series solution which can be controlled by the convergence control parameter $c_0$ if the homotopy is constructed with respect to the generalized Taylor series. The result for the asymptotic series found for the toy model can be readily generalized to a simple FT where the asymptotic series solution can be controlled even close to a 2nd order phase transition.\\

{\it Acknowledgements} -- The authors would like to thank D. H\"ugel, A. Toschi and the participants of the \textit{Diagrammatic Monte Carlo Workshop} held at the Flatiron Institute, June 2017 for fruitful discussions and valuable input. This work is supported by FP7/ERC Starting Grant No. 30689, FP7/ERC Consolidator Grant No. 771891, and the Nanosystems Initiative Munich (NIM).

%
%

%
%
\newpage
%
%
\section{Supplemental Material: Analytic structure of the universal functional $\Gamma[G]$ - Convergent HAM series solution}
\begin{figure}[b]
\centering
\includegraphics[width=0.9\linewidth]{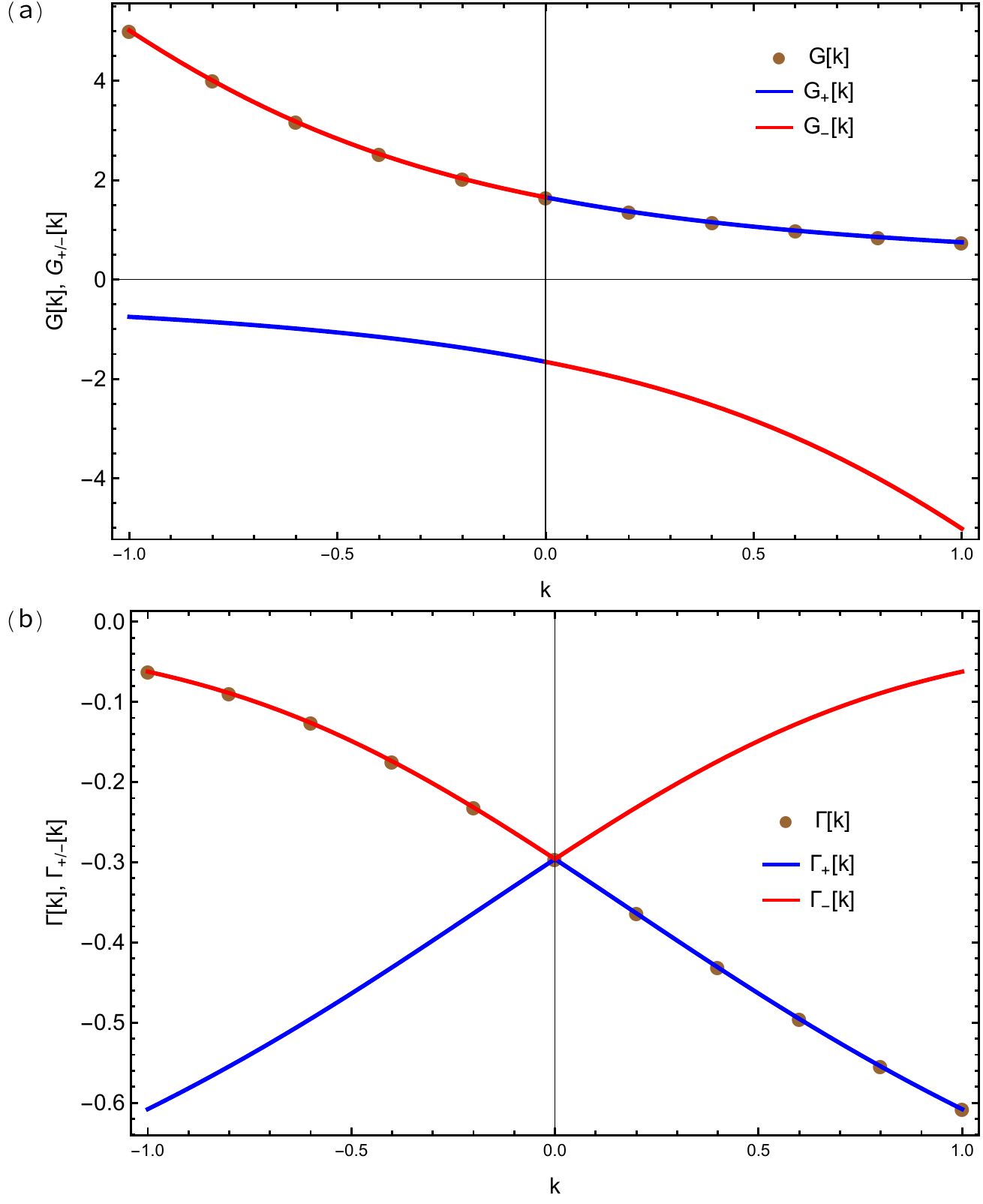}
\caption{The two solutions for $G(\Gamma)_{\pm}[k]$ of (\ref{AppCoupledink}) on the real axis for $\lambda=1$. The physical solution $G(\Gamma)[k]$ corresponds to the solution $G(\Gamma)_+[k]$ for $k>0$ and $G(\Gamma)_-[k]$ for $k<0$ cf. (\ref{Appanalytic}).
}
\label{fig:TwoSolutions}
\end{figure}
In order to obtain the analytic structure of $\Gamma[G]$ near $G=0$ we first consider the analytic solution to the integral
\begin{equation}
Z = \int \!\! \text{d}\phi \; e^{ -S[\phi] } \text{ with } S[\phi] = \frac{1}{2} k \phi^2  + \frac{\lambda}{4!} \phi^4.
\label{appIntegral}
\end{equation}
The analytic solution to this integral for $k \in \mathbb{C}$ with fixed $\lambda \in \mathbb{R}^+$ is given as
\begin{equation}
\begin{split}
Z[k]  & =  e^{\frac{3k^2}{\lambda}} \sqrt{\pm \frac{k}{\lambda}} \\ 
& \times \begin{cases}
\sqrt{6} K_{\frac{1}{4}}\left( \frac{3k^2}{\lambda} \right) & \text{Re}(k) > 0 \\
\sqrt{2} \left( I_{\frac{1}{4}}\left( \frac{3k^2}{\lambda} \right) + I_{-\frac{1}{4}}\left( \frac{3k^2}{\lambda} \right) \right)& \text{Re}(k) < 0.  
\end{cases}
\end{split}
\label{Appanalytic}
\end{equation} 
\begin{figure}[t]
\centering
\includegraphics[width=0.9\linewidth]{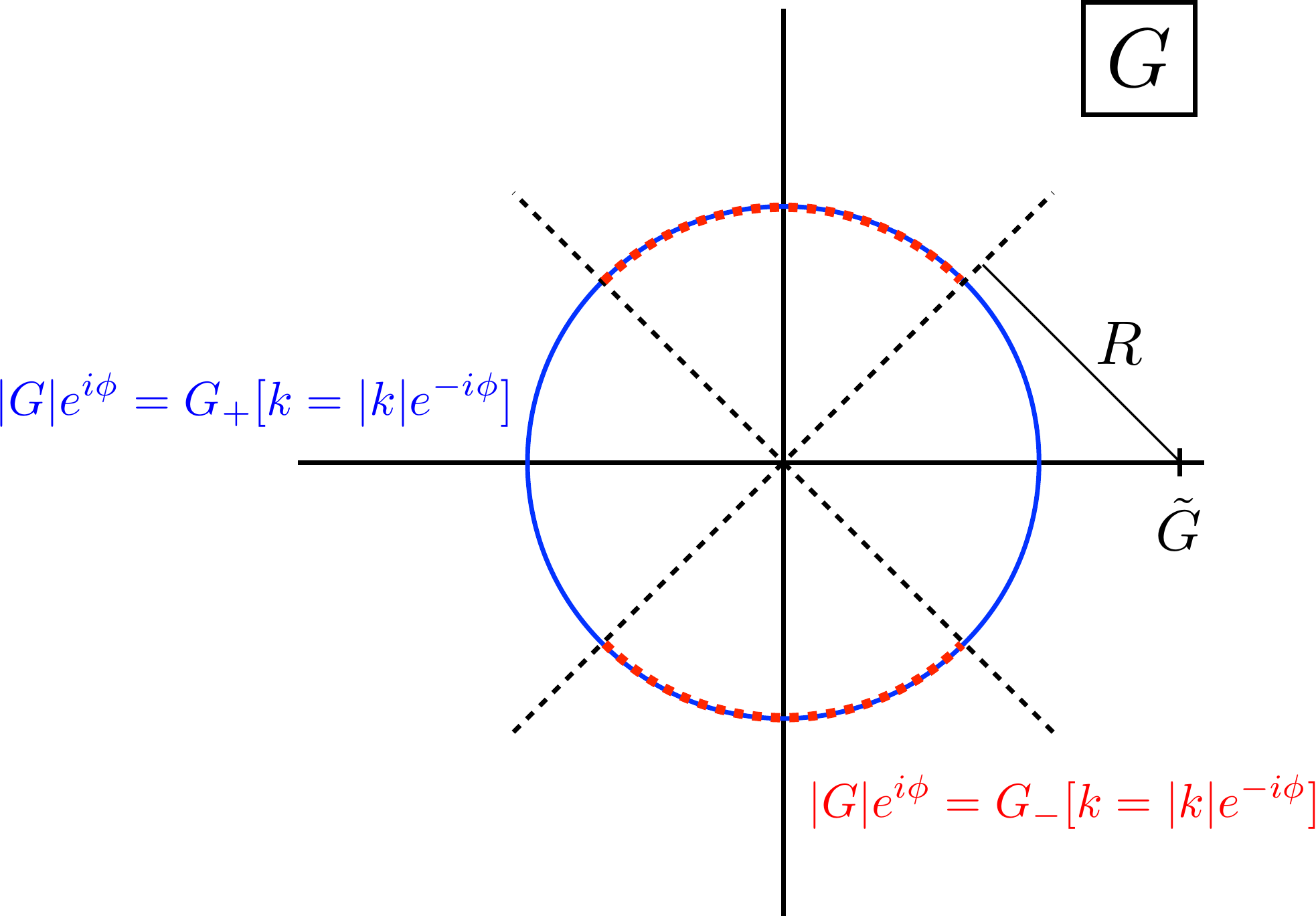}
\caption{The elements of the images of $G_{\pm}[k]$ onto the circle $|G| e^{i\phi}$ for $ |G| \rightarrow 0$. For $\phi \in [\frac{\pi}{4},\frac{3\pi}{4}] $ and $[\frac{5\pi}{4}, \frac{7\pi}{4}]$ $G_{\pm}$ map onto the same $G$. Therefore, $\Gamma[G]$ has to be single valued on $\phi \in [-\frac{\pi}{4},\frac{\pi}{4}] $ and $[\frac{3\pi}{4}, \frac{5\pi}{4}]$ and multi-valued on $\phi \in [\frac{\pi}{4},\frac{3\pi}{4}] $ and $[\frac{5\pi}{4}, \frac{7\pi}{4}]$. The Taylor series of $\Gamma[G]$ around $\tilde{G} \in \mathbb{R}^+$ has the convergence radius $R=\sin(\frac{\pi}{4}) \tilde{G}$. }
\label{fig:BranchCuts}
\end{figure}
Here $I_n(z)$/$K_n(z)$ are the modified Bessel functions of the first/second kind. The integral (\ref{appIntegral}) can be thought of as an integral representation of the function $Z$ defined in (ref{Appanalytic}). The same holds for the 2-point correlation function $G[k]$ which we do not show here. The important point is that both $Z[k]$ and $G[k]$ are piecewise defined functions in $k$ which can be represented by a single integral expression. There exists also a differential equation equivalent to the integral representation of the function $Z[k]$ or $G[k]$.  For $G[k]$ this is the coupled system of differential equations,
\begin{equation}
\begin{split}
&G_{\pm}[k]^{-1} - k = \frac{\lambda}{2} G_{\pm}[k] + \frac{\lambda}{6} G_{\pm}[k]^3 \Gamma_{\pm}^{(4)}[k] \\
&\Gamma_{\pm}[k] = -\lambda - \frac{3\lambda}{2} G_{\pm}[k]^2 \Gamma_{\pm}[k] - \frac{\lambda}{2} G[k]_{\pm}^4 \left[\Gamma_{\pm}^{(4)}\right]^2 \\
& \hspace{15mm} + \frac{\lambda}{6} G_{\pm}[k] \frac{ \text{d} \Gamma_{\pm}^{(4)}}{ \text{d} k }.
\label{AppCoupledink}
\end{split}
\end{equation}
\begin{figure}[b]
\centering
\includegraphics[width=0.9\linewidth]{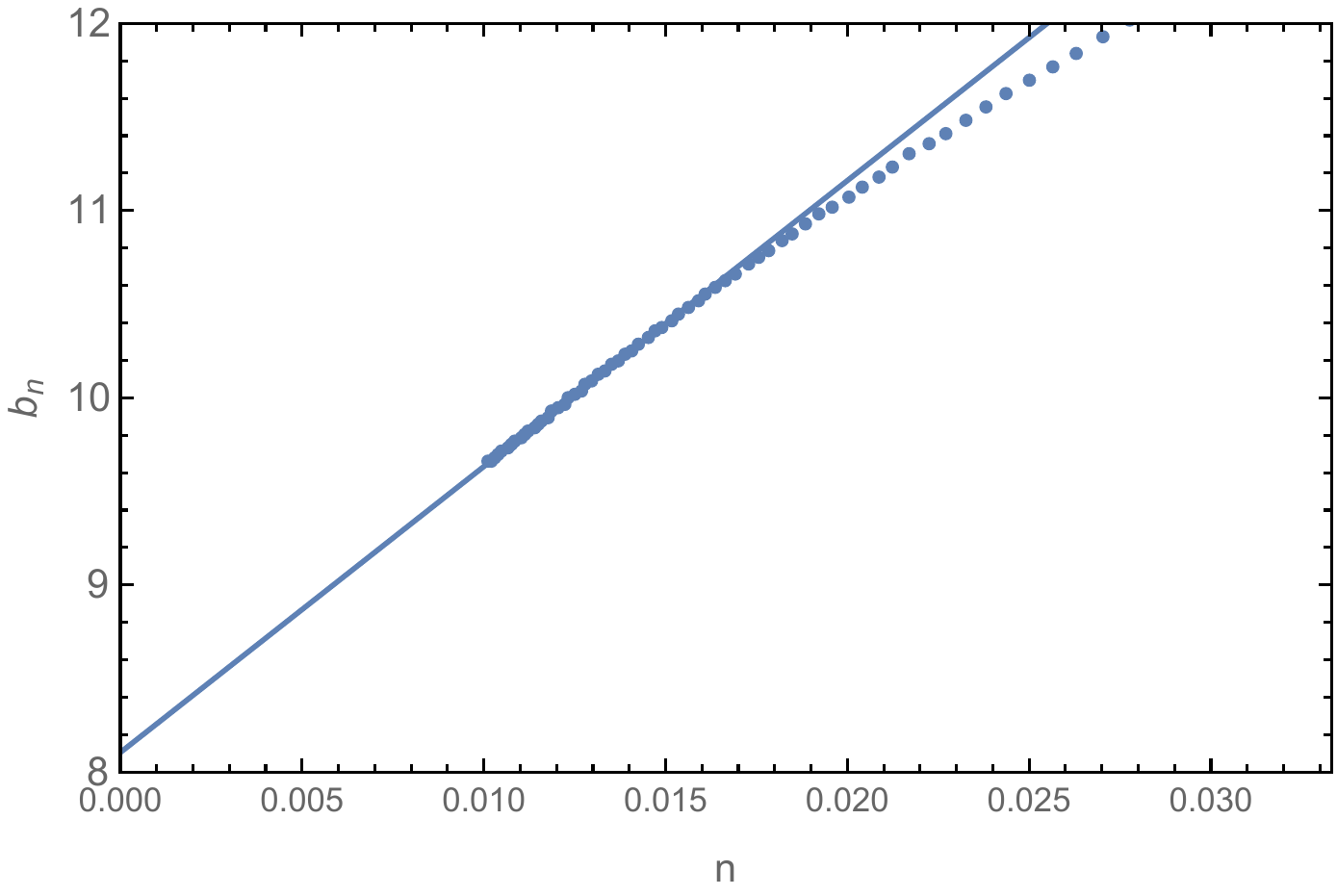}
\caption{The linear extrapolation of the large order asymptotic $b^2_n = \frac{c_{n+1}c_{n-1}-c^2_n}{c_n c_{n-2} - c^2_{n-1}}$ yields a convergence radius $R \approx \frac{1}{8.105} = 0.123$ which can be compared to the expected convergence radius of $R = \sin(\frac{\pi}{4}) 0.172 $ = 0.122. Here $\tilde{G} = G_+[k=5] = 0.172$ has been considered.}
\label{fig:LargeOrderPowerSeries}
\end{figure}
There are two linearly independent solutions to this equation for $G$ and, consequently, also for $\Gamma$ denoted as $G_{\pm}[k]$ and $\Gamma_{\pm}[k]$ respectively. 
While (\ref{appIntegral}) is single-valued, the solution to the system of equations (\ref{AppCoupledink}) in principle leads to two independent solutions. They are shown for $G_{\pm}[k]$, $\Gamma_{\pm}[k]$ on the real axis in Fig.\,\ref{fig:TwoSolutions}. We will show in the following that it is not the non-analytic behavior of $G[k]$ around $k=0$ which determines the analytic structure of $\Gamma[G]$ near $G=0$ but it is the existence of the two independent solutions to (\ref{AppCoupledink}).\\
\begin{figure}[b]
\centering
\includegraphics[width=0.9\linewidth]{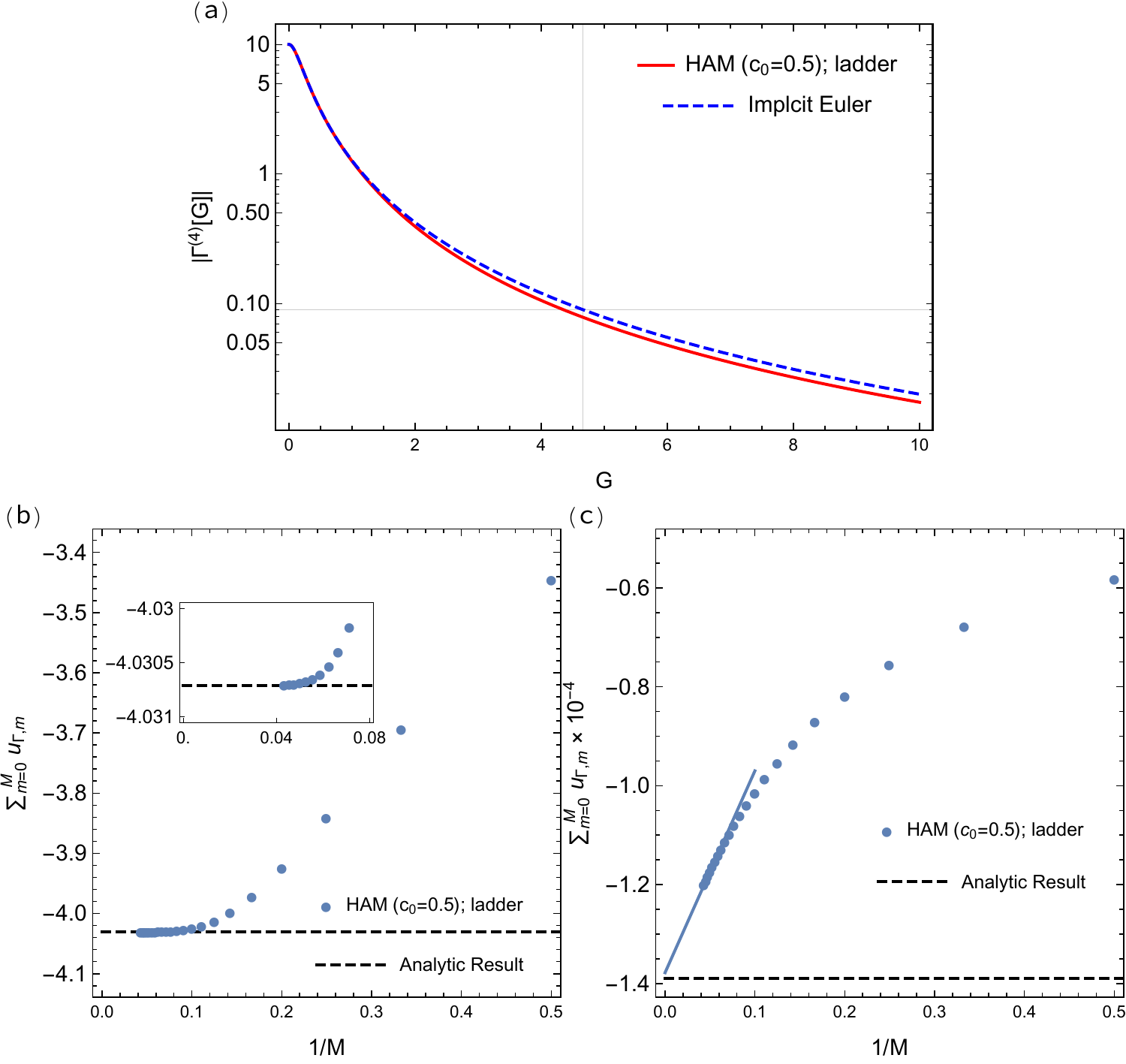}
\caption{ The approximation to the HAM series solution $\Gamma[G] = \lim_{m \rightarrow \infty} \sum_m^M u_{\Gamma,m}$ with $M=22$. It gives not only locally a convergent result as with a generalized Taylor series but also globally. The truncation of the series solution at $M=22$ already gives decent results for large $G$ without extrapolating the series solution. The thin grid lines show the analytic result for the evaluation of $\Gamma[G]$. The convergence of the HAM series solution is demonstrated in (b) and (c). Without tuning the convergence control parameter $c_0$, convergence can be observed both for small ( (b) $G[k=1] \approx 0.4$ ) and large ( (c) $G[k=-100] \approx 12$) $G$.}
\label{fig:GammaGGlobal}
\end{figure}
Formally, $\Gamma[G]$ can be obtained from $\Gamma_{\pm}[k]$ by inverting the relation $G = G_{\pm}[k]$. There are two independent solutions which for some $k$ satisfy $G = G_+[k_1] = G_-[k_2]$. This leads to two branches for $\Gamma[G]$ where $\Gamma[G]$ must evaluate to $\Gamma_+^{(4)}[k_1]$ on the first branch and to $\Gamma_-^{(4)}[k_2]$ on the second branch. In order to obtain the analytic structure of $\Gamma[G]$ near $|G|=0$ it is necessary to study the intersection of the images of $G_{\pm}$ for elements which satisfy $|G| \to 0$. The complete circle in the complex $G$ plane $ |G| e^{i\phi}$, $\phi \in [0,2\pi]$ for $|G| \rightarrow 0$ in Fig.\,\ref{fig:BranchCuts} is included in the image of $G_{+}[k]$ and can be obtained by the parametrization $k = |k| e^{ - i \phi}$ where $|k|\rightarrow \infty$. On the other hand, the image of $G_-[k]$ includes two segments of the circle $ |G| e^{i\phi}$ namely with $\phi \in [\frac{\pi}{4}, \frac{3 \pi}{4}]$ and $\phi \in [\frac{5\pi}{4}, \frac{7 \pi}{4}]$ and can also be obtained by the parametrization $k = |k| e^{ - i \phi}$. Therefore there are branch cuts starting from $G=0$ and cutting the complex $G$ plane in the $\pm \frac{\pi}{4}$ direction.\\
According to the above result a power series solution around $\tilde{G} \in \mathbb{R^+}$, should have convergence radius $R = \sin(\frac{\pi}{4}) \tilde{G}$. 
This convergence radius can be numerically determined by considering the coefficients $c_n$ in the power series $\Gamma[G] = \sum_n c_n (G-\tilde{G})^n$. The power series can be obtained by using the analytically known result for $\Gamma_+^{(4)}[G_+[k]]$ on the real axis i.e. $c_0 = \Gamma[\tilde{G}]$.
As shown in Fig.\,\ref{fig:LargeOrderPowerSeries} the numerically obtained convergence radius from the large order behavior of the coefficients $c_n$ agrees with the expected convergence radius.\\
Moreover, we numerically find a convergent HAM series solution if we are considering a linear operator $\mathcal{L}$ such that the HAM does not yield a series solution which is a generalized Taylor series around the non-interacting limit $G=0$.
With the simple choice $\mathcal{L_{\text{ladder}}}\left[\Gamma[G]\right] = \Gamma[G] + \lambda + \frac{3 \lambda}{2} G^2 \Gamma[G]$ we obtain the results shown in Fig.\,\ref{fig:GammaGGlobal}(a). The linear operator is the ladder approximation to the functional $\Gamma[G]$ and therefore the HAM series solution is an expansion around this ladder approximation.
With the linear operator $\mathcal{L}_{\text{ladder}}$ the HAM series solution gives not only a locally convergent solution to the functional $\Gamma[G]$ but it gives a globally convergent solution. This is demonstrated in Figs.\,\ref{fig:GammaGGlobal}(b) and (c). The HAM series solution is converging for arbitrary large $G$ without tuning the convergence control parameter $c_0$.
%
%
%
\end{document}